\documentclass[10pt,a4paper,twocolumn,showpacs,superscriptaddress,amsmath,amssymb,aps,prl,preprintnumbers]{revtex4-2}
\usepackage{bbm}
\usepackage{dsfont}
\usepackage{verbatim}
\usepackage{graphicx}
\usepackage{braket}
\usepackage[dvipsnames, svgnames, x11names]{xcolor}
\usepackage{array}
\usepackage{bibunits}
\usepackage{mathrsfs}
\usepackage{newtxmath}
\usepackage{dcolumn}
\usepackage{bm}
\usepackage{lineno}
\usepackage{hyperref}
\hypersetup{hypertex=true,
	colorlinks=true,
	linkcolor=blue,
	anchorcolor=blue,
	citecolor=blue}

\begin{document}
\title{Quantum trajectory of a spin oscillator, continuous retrodiction beyond the Heisenberg uncertainty relation}
\author{Peiying Li}%
\altaffiliation{These authors contributed equally to this work.}
\affiliation{State Key Laboratory of Quantum Optics Technologies and Devices, Institute of Laser Spectroscopy, Shanxi University, Taiyuan, Shanxi 030006, China}%
\affiliation{Collaborative Innovation Center of Extreme Optics, Shanxi University, Taiyuan 030006, China}
\author{Junlei Duan}
\altaffiliation{These authors contributed equally to this work.}
\email{20110190074@fudan.edu.cn}
\affiliation{State Key Laboratory of Surface Physics, Department of Physics and Key Laboratory of Micro and Nano Photonic Structures (Ministry of Education), Fudan University, Shanghai 200433, China}%
\author{Suotang Jia}%
\affiliation{State Key Laboratory of Quantum Optics Technologies and Devices, Institute of Laser Spectroscopy, Shanxi University, Taiyuan, Shanxi 030006, China}%
\affiliation{Collaborative Innovation Center of Extreme Optics, Shanxi University, Taiyuan 030006, China}
\author{Klaus M\o lmer}%
\email{klaus.molmer@nbi.ku.dk} 
\affiliation{Niels Bohr Institute, University of Copenhagen, Jagtvej 155A, DK 2200 Copenhagen, Denmark}%
\author{Yanhong Xiao}%
\email{yxiao@fudan.edu.cn}
\affiliation{State Key Laboratory of Surface Physics, Department of Physics and Key Laboratory of Micro and Nano Photonic Structures (Ministry of Education), Fudan University, Shanghai 200433, China}%
\affiliation{State Key Laboratory of Quantum Optics Technologies and Devices, Institute of Laser Spectroscopy, Shanxi University, Taiyuan, Shanxi 030006, China}%
\affiliation{Collaborative Innovation Center of Extreme Optics, Shanxi University, Taiyuan 030006, China}%
%
\begin{abstract}
	The Heisenberg uncertainty relation in quantum mechanics limits the precision of predicting the outcome of measurements of any non-commuting observables. However, this restriction does not apply to retrodictive guessing of the outcomes of past measurements. Here, we experimentally demonstrate this using a spin oscillator consisting of $ 5.17 \times 10^{10}$ rubidium atoms. Through weak continuous measurements, we monitor canonical spin observables $\hat{x}_{\mathrm{A}}$ and $\hat{p}_{\mathrm{A}}$ with $[\hat{x}_{\mathrm{A}},\hat{p}_{\mathrm{A}}] = i\hbar$. By combining information obtained before and after time $t$, we retrodict the outcomes of measurements of $\hat{x}_{\mathrm{A}}$ and $\hat{p}_{\mathrm{A}}$ at time $t$ with variances that remain, concurrently over 1.4 hours, $\sim$ 1.01 dB  below the bound set by the Heisenberg uncertainty relation. We also infer and demonstrate the ability to retrodict, with the same precision, the measurement outcomes for arbitrary quadratures of the spin oscillator. This brings interesting prospects for further research on quantum measurements, and for quantum-enhanced sensing that combines trajectory and entanglement strategies.	
\end{abstract}
\maketitle
	
Articulated first by Werner Heisenberg in 1927, the Heisenberg uncertainty relation (HUR) asserts a fundamental limitation on predicting the outcome of measurements of two non-commuting observables \cite{heisenberg1927anschaulichen}. A well-known example is that the more precisely one can guess the outcome of measuring a particle's position, the less precisely one can guess the outcome of a momentum measurement, and vice versa. For canonically conjugate observables $\hat{x}$ and $\hat{p}$ with $[\hat{x},\hat{p}] = i\hbar$, the HUR can be formally expressed as $\Delta x \cdot \Delta p \geq \hbar/2$ \cite{robertson1929uncertainty}. Over the past century, this seminal theoretical framework has not only served as a critical component in the interpretation of quantum mechanics but has also revolutionized our conceptual understanding of measurement limitations at the quantum scale. A profound consequence of HUR is the so-called standard quantum limit (SQL) in precision measurement, which defines the best precision achievable using classical resources such as coherent states. State-of-the-art sensors, including LIGO \cite{aasi2013enhanced}, optical lattice clocks \cite{pedrozo2020entanglement,robinson2024direct}, atomic interferometers \cite{greve2022entanglement,malia2022distributed}, and optical magnetometers \cite{sewell2012magnetic,bao2020spin,zheng2023entanglement}, operate near SQL-limited regimes, and quantum correlations and entanglement have been exploited to go beyond the SQL \cite{clerk2010introduction,pezze2018quantum}.

A complementary aspect of relevance to practical precision measurements is the disturbance on quantum states that are subject to measurement, a phenomenon referred to as quantum back-action (QBA). A variety of quantum measurement strategies have been proposed or demonstrated to circumvent or  mitigate the measurement constraints arising from the QBA. These approaches include: `coherent quantum-noise cancellation' \cite{tsang2010coherent}, `quantum-mechanics-free systems' \cite{tsang2012evading}, `negative mass' schemes \cite{ockeloen2018stabilized,thomas2021entanglement,julsgaard2001experimental,mercierdelepinay2021quantum,novikovHybridQuantumNetwork2025}, `planar squeezed states' \cite{colangelo2017simultaneous}, `quantum memory' \cite{berta2010uncertainty,li2011experimental,prevedel2011experimental}, `sequential modular encoding' \cite{asadian2014probing,fluhmann2018sequential,valahu2025quantum}, and `spin-nematic squeezed state' \cite{gross2011atomic,hamley2012spin,cao2025joint}. The key idea of these methods is to strategically measure appropriately designed commuting observables to indirectly infer non-commuting observables, or by redirecting quantum back-action noise into unmeasured degrees of freedom.

\begin{figure*}[t]
	\centering
	\includegraphics[width= 7 in]{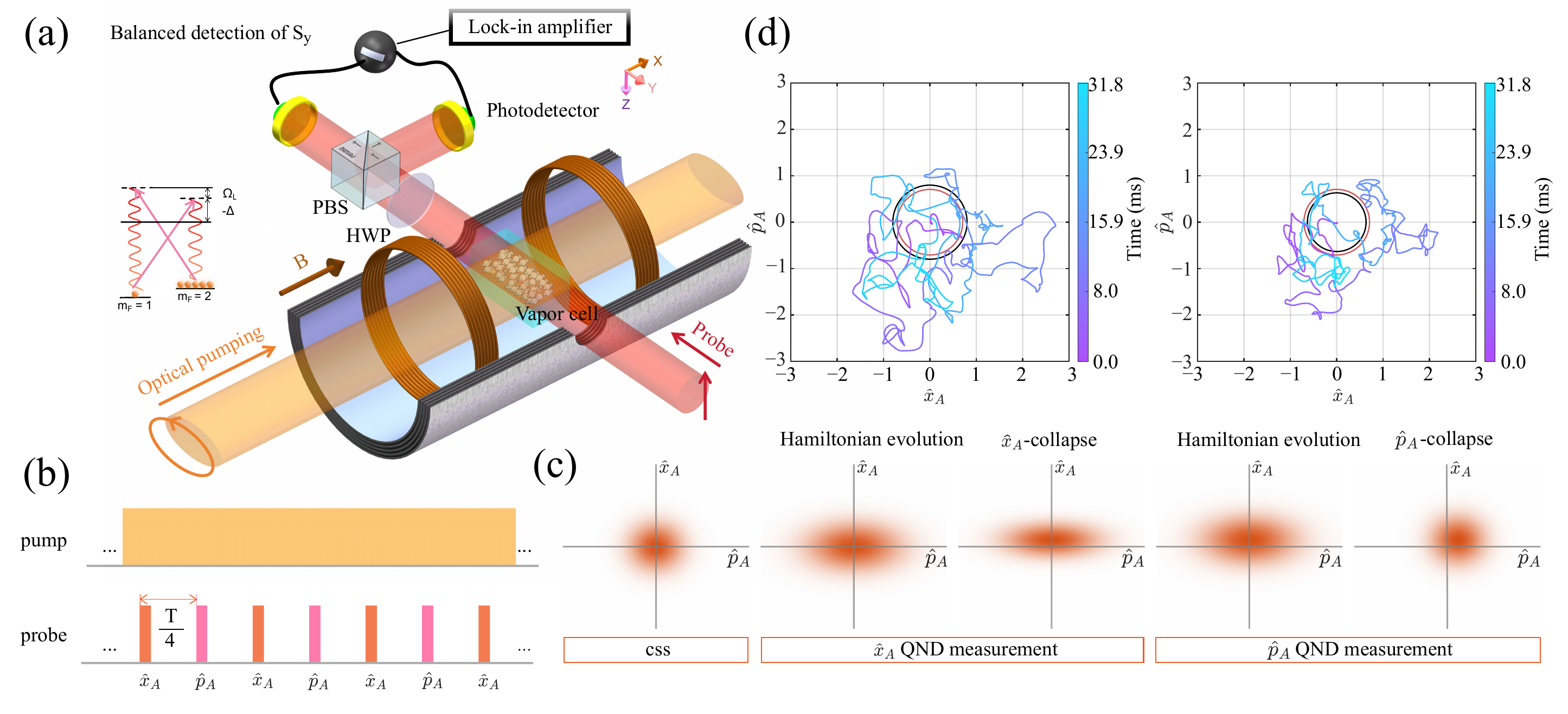}	
	\caption{{\bf Experimental setup and principles.} 
		{\bf (a) Experimental setup.} {A paraffin-coated $^{87}$Rb vapor cell is placed inside a five-layer magnetic shield and maintained at $58^\circ\mathrm{C}$. A static bias magnetic field is applied along the $x$ axis. Continuous optical pumping prepares a highly polarized ensemble, establishing a macroscopic collective spin $J_x$. A linearly polarized probe beam, blue-detuned by $4.5~\mathrm{GHz}$ from the $\ket{5S_{1/2},F=2}\rightarrow\ket{5P_{3/2},F'=3}$ transition, passes through the cell. The Faraday rotation is detected with a balanced polarimeter, enabling continuous monitoring of the transverse spin components $J_y$ and $J_z$, which precess about the $x$ axis at the Larmor frequency $\Omega_L=2\pi\times500.35~\mathrm{kHz}$.}
        {\bf (b) Experimental pulse sequence.} {The optical pumping light is continuously on during the experiment, while the probe light is modulated at four times the Larmor frequency for stroboscopic detection, enabling effectively simultaneous measurement of both $\hat{x}$ and $\hat{p}$ in the rotating frame.}
        {\bf (c) Illustration of the uncertainty cloud evolution.} {The figure shows the evolution of the Wigner function in phase space during the measurement process. When a QND measurement is performed on the $x$ quadrature, the uncertainty in the $p$ quadrature increases due to the measurement back-action, and vice versa.}        
        {\bf (d) Inferred phase-space trajectory of the spin oscillator.}{ The left panel shows a single-shot phase-space trajectory \((x_{\rm A}(t),p_{\rm A}(t))\) of the quadrature expectation values according to the density matrix \(\rho(t)\) conditioned on the measurements until time \(t\). The right panel shows the corresponding retrodicted trajectory \((x_{\rm A}^{\rm r}(t),p_{\rm A}^{\rm r}(t))\) obtained by combining the density matrix \(\rho(t)\) with the matrix \(E(t)\) conditioned on the measurements obtained after time \(t\). All quantities are displayed in oscillator units, and the line color indicates time during the measurement sequence. The black circles represent the expected standard deviations of quadrature measurements around the trajectory expectation values, which is larger (smaller) for prediction (retrodiction) than the standard quantum limit, shown with the red circle in the two panels.}}
	\label{Fig:1}			
\end{figure*}

In this article, we pursue quantum trajectory theory, which enables enhanced estimation of system dynamics and more precise characterization of physical parameters and time-dependent perturbations \cite{gammelmark2013past, zhang2017prediction, tsang2009time, guevara2015quantum}. In this framework, the density matrix $\rho (t)$ is governed by the dynamics of the system and measurement record up to time $t$. $\rho (t)$, in turn, yields the probabilities of the possible outcomes $m$ of the subsequent measurement on the system, $P(m)=$Tr$(\Omega_m\rho(t)\Omega_m^{\dagger})$, where the operators $\Omega_m$ generalize the Born rule beyond its application to projective measurements \cite{bergou2010discrimination}. The outcome of the measurement, in turn, updates the density matrix, $\rho(t) \rightarrow \Omega_m\rho(t)\Omega_m^{\dagger}/P(m)$ and, hereby, conditions the probability of subsequent measurements on the outcome $m$ at time $t$. Knowledge of the measurement outcomes obtained during the future interval $(t, T]$ therefore modifies in a Bayesian manner, the likelihood of the different outcomes at time $t$. This knowledge contributes, via a matrix $E(t)$, to the {\it retrodictive} conditional probability $P^{ret}(m)\propto\mathrm{Tr}(\Omega_m\rho(t)\Omega_m^{\dagger}E(t))$ \cite{gammelmark2013past}. 
For weak measurements, this yields results in agreement with Wiseman's concept of a smoothed quantum state $\rho_S \propto \rho_F(t)\rho_R(t)$ \cite{guevara2015quantum}, and it is also in agreement with the weak values associated with pre- and post-selected pure quantum states \cite{aharonov1988result}, and with the ABL rule for strong, projective measurements in the same setting \cite{aharonov1964time}. 

The retrodiction and smoothing formulas are special cases of a hybrid classical-quantum theory, pioneered by Tsang \cite{tsang2009time,tsang2009optimal} to estimate classical Markov processes coupled to a quantum system that is subject to continuous measurements. We are not concerned with classical stochastic processes affecting our quantum system, but with how our ability to guess the outcome of quantum measurements is fundamentally modified by knowledge of earlier and later measurement results. The expression for $P^{ret}(m)$ has been confirmed experimentally by application of the formalism to cavity QED \cite{rybarczyk2015forward}, superconducting qubits \cite{tan2015prediction}, and to atomic ensembles \cite{bao2020spin}. The retrodiction protocol has been used to infer the outcome of measurements of non-commuting quadratures of a mechanical oscillator \cite{rossi2019observing} and of a collective spin oscillator \cite{bao2020retrodiction}. While it was experimentally demonstrated in  \cite{bao2020retrodiction} that the conventional Heisenberg Uncertainty Relation does not apply in the case of retrodiction because the outcome of measurements of position and momentum could be separately inferred from earlier and later measurements, their linear combinations were in the same study subject to larger uncertainties. Simultaneous sub-SQL retrodiction of arbitrary linear combinations of position and momentum quadratures has been analyzed theoretically \cite{khanahmadi2021guessing} but has not yet been demonstrated experimentally.

Here, we experimentally demonstrate simultaneous estimation of non-commuting variables beyond the Heisenberg uncertainty bound and verify the results using weak measurement. The $\hat{x}_A$ and $\hat{p}_A$ components of the oscillator are continuously monitored, allowing us to reconstruct the quantum trajectory. By jointly considering the forward evolving density matrix and the backward evolving effect matrix, we show that the measurement uncertainty can be reduced to the level of $-1.01 \pm 0.15$\: dB compared to the limit imposed by the HUR. Remarkably, the conditional variances of both $\hat{x}$ and $\hat{p}$, and of arbitrary quadratures in phase space, can simultaneously fall below the SQL. The key finding is that in simultaneous estimation of non-commuting variables, the below-SQL variances are only obtainable in double-sided estimation (i.e., conditioned on measurement results both before and after), and that for single-sided estimation (i.e., conditioned only on results before) the variances are at best at the SQL level. Our approach is applicable to both continuous and pulsed measurements, and can be extended to the simultaneous estimation of position and momentum in mechanical resonators or other spin systems, where the estimation precision of both non-commuting observables could lie below the SQL.

The core of the experiment is an atomic vapor of $^{87}$Rb with an atom number $N_A \approx 5.17 \times 10^{10}$ , confined within an anti-relaxation-coated glass cell. The cell is maintained at $58^\circ\mathrm{C}$ using temperature-controlled airflow inside a five-layer magnetic shield, which minimizes electromagnetic noise and preserves long spin-coherence times (Fig.\:\ref{Fig:1}a). The collective spin state of atoms is defined as $\boldsymbol{\hat{J}}= \sum_k \boldsymbol{\hat{j}}^k$, where $k$ labels individual atomic spins. Continuous optical pumping prepares the atomic ensemble to the Zeeman sublevel $\ket{5 S_{1/2}, F=2, m_F = 2}$ with a degree of spin polarization of $96.91\%$ \cite{duan2025concurrent}. The highly polarized atomic spin can be treated as a spin oscillator under the Holstein-Primakoff transformation \cite{vasilakis2015generation}, and the canonical position and momentum operators are $\hat{x}_{\mathrm{A}}=\hat{J}_{\mathrm{y}} / \sqrt{\left|\left\langle J_{\mathrm{x}}\right\rangle\right|}$ and $\hat{p}_{\mathrm{A}}=\hat{J}_{\mathrm{z}} / \sqrt{\left|\left\langle J_{\mathrm{x}}\right\rangle\right|}$. The two canonical operators obey $[\hat{x}_{\mathrm{A}},\hat{p}_{\mathrm{A}}] = i$ (Hereafter, $\hbar = 1$), and the HUR is $\Delta x \cdot \Delta p \geq 1/2$. A static magnetic field of 0.71 G is applied along the $x$-direction, and $\hat{x}_{\mathrm{A}}$ and $\hat{p}_{\mathrm{A}}$ oscillate in the lab frame at the Larmor frequency $\Omega_L$. The position and momentum variables in the laboratory frame (with the primes) are $\hat{x}_{\mathrm A}^{\prime}=\hat{x}_{\mathrm A}\cos(\Omega_L t) +\hat{p}_{\mathrm A}\sin(\Omega_L t)$ and $\hat{p}_{\mathrm A}^{\prime}
=-\hat{x}_{\mathrm A}\sin(\Omega_L t)+\hat{p}_{\mathrm A}\cos(\Omega_L t)$. 

\begin{figure*}[t]
	\centering
	\includegraphics[width=7in]{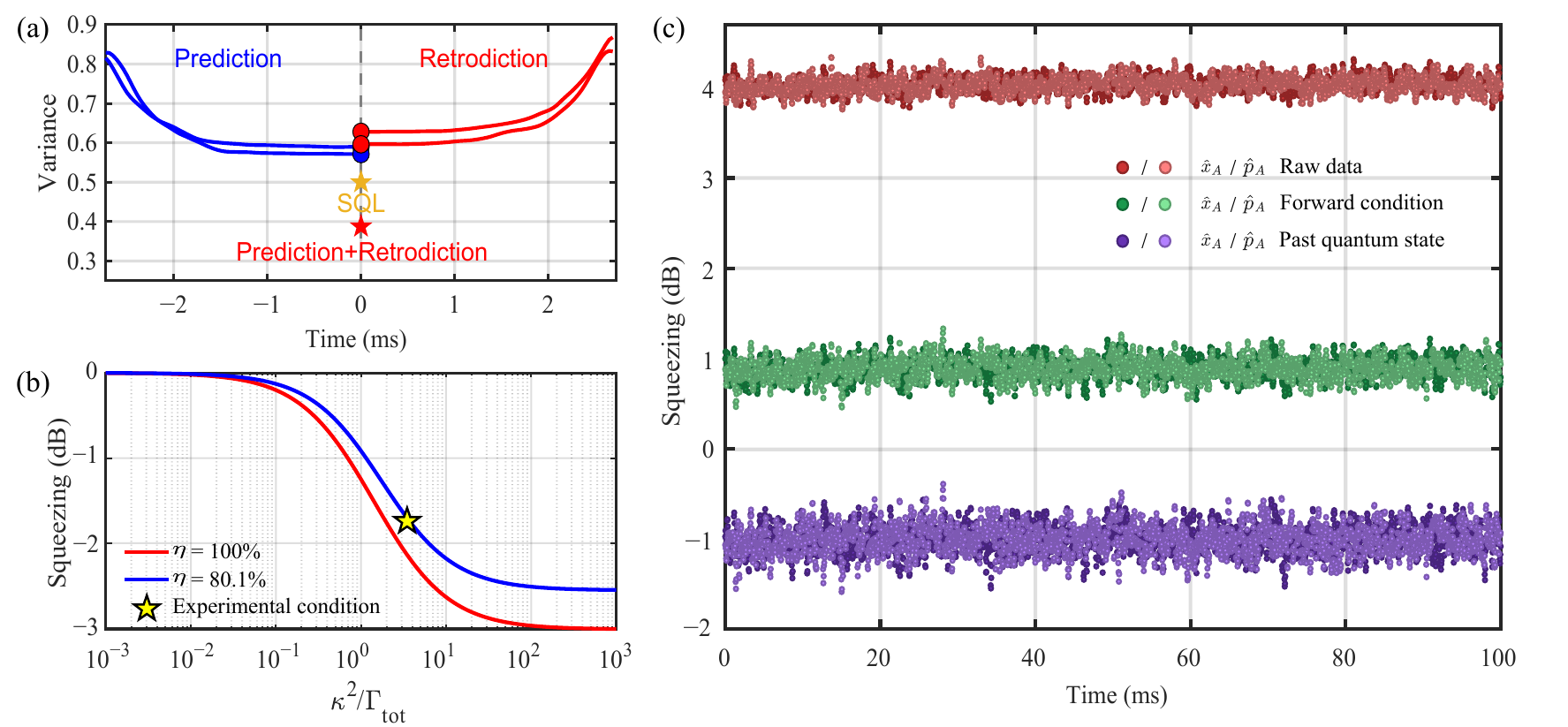}	
	\caption{{\bf Simultaneous squeezing of the canonical $\hat{x}$ and $\hat{p}$ spin quadratures.} 
		{\bf (a) Time evolution of the variance.} {The blue curves show the variances of the probability distributions of measurement outcomes conditioned on the measurement record until $t=0$. The two solid blue lines correspond to the time evolution of the conditional variances of the quadratures $\hat{x}$  and $\hat{p}$, respectively. In theory, these variances should overlap completely due to the symmetric nature of the system, but slight deviations arise from unavoidable experimental imperfections. Similarly, the red curves represent the variances for the likelihood function assigned to prior measurements conditioned by the later data.  The yellow asterisk denotes the variance associated with the standard quantum limit (SQL). By combining the information in the past and the future of $t=0$, we retrodict the measurement outcome at  $t=0$ with both variances reduced to 0.38, which lies below the standard quantum limit, as illustrated by the red asterisk. }
		{\bf (b) Theoretical curve for squeezing.}{ The plot shows how the theoretically predicted squeezing level depends on the ratio of the coupling strength $ \kappa^2$ to the total damping rate $ \Gamma_{\mathrm{tot}}$, where symmetric coupling is assumed: $ \kappa=\kappa_x = \kappa_p$.}
		{\bf (c) Steady-state simultaneous squeezing of the quadratures $\hat{x}$ and $\hat{p}$.} {Without conditioning, the quadratures $\hat{x}$ and $\hat{p}$ exhibit fluctuations with a variance $4.03~\mathrm{dB}$ (red and light red) above the SQL. The measured data fluctuate with a variance of  $0.88~\mathrm{dB}$ above the SQL around the value predicted by prior measurements (green and light green), but with a variance of $1.01~\mathrm{dB}$ below the SQL around the value inferred by retrodiction (purple and light purple). The scatter plot shows that the reduction in conditional variance is stable over durations exceeding $100\:\mathrm{ms}$, with each data point derived from 50000 repeated measurements.}
		}
	\label{Fig:2}			
\end{figure*}

Continuous QND measurements on the spin oscillator's quadratures $\hat{x}_{\mathrm{A}}$ and $\hat{p}_{\mathrm{A}}$ are performed via Faraday rotation using a far-detuned probe laser. The interaction is governed by a typical QND Hamiltonian, $\hat{H} = \kappa \hat{p}_{\mathrm{A}}^{\prime} \hat{p}_{\mathrm{L}}$, where $\hat{p}_{\mathrm{A}}^{\prime}$ is undisturbed during the measurement \cite{hammerer2010quantum}. The canonical light operators (Stokes vector) are defined as $\hat{x}_{\mathrm{L}}=\hat{S}_{\mathrm{y}} / \sqrt{\left|\left\langle S_{\mathrm{x}}\right\rangle\right|}$ and $\hat{p}_{\mathrm{L}}=\hat{S}_{\mathrm{z}} / \sqrt{\left|\left\langle S_{\mathrm{x}}\right\rangle\right|}$. During spin precession in the laboratory frame, $\hat{x}_{\mathrm{A}}$ and $\hat{p}_{\mathrm{A}}$ are measured sequentially. We use a stroboscopic technique to restrict the measured observables $\hat{p}_{\mathrm{A}}^{\prime}$ to $\hat{x}_{\mathrm{A}}$ and $\hat{p}_{\mathrm{A}}$ (Fig.\:\ref{Fig:1}b). The probe light is modulated at four times the Larmor frequency to enable intertwining and alternating measurements of $\hat{x}_{\mathrm{A}}$ and $\hat{p}_{\mathrm{A}}$, which is the key design of this experiment. The two quadratures are detected through the $\hat{x}_{\mathrm{L}}$ of the output light and can be easily separated as they are $\pi/2$ out of phase. During the probing process in Fig.\:\ref{Fig:1}c, while measuring $\hat{p}_{\mathrm{A}}^{\prime} = \hat{x}_{\mathrm{A}}$, $\rm{Var}\{\hat{p}_{\mathrm{A}}\}$ first increases due to the atom-light  interaction and $\rm{Var}\{\hat{x}_{\mathrm{A}}\}$ decreases as $\hat{x}_{\mathrm{L}}$ is detected. The next pulse addresses the orthogonal quadrature $\hat{p}_{\mathrm{A}}^{\prime} = \hat{p}_{\mathrm A}$, reducing $\mathrm{Var}\{\hat{p}_{\mathrm A}\}$ through information gained from the detected light while adding quantum back-action noise to $\hat{x}_{\mathrm A}$. Each measurement pulse here is short in time ($0.1\:\mu \mathrm{s}$) and very weak, and when we integrate the measurements over a period of time ($37.3\:\mu \mathrm{s}$) we effectively implement simultaneous and continuous measurements of $\hat{x}_{\mathrm{A}}$ and $\hat{p}_{\mathrm{A}}$. In the experiment here, the measurement rate is $\kappa^2 \eta = 2\pi \times 1002\:{\rm Hz}$, where the total quantum efficiency $\eta = 80.1\:\%$, mainly limited by the transmittance of the paraffin coating and the quantum efficiency of photodetectors. 

While the measurement continues, we obtain the measurement records, $\mathbf{R}=\{\mathbf{R}_x,\mathbf{R}_p\}$. In the rotating frame of the spin oscillator and by approximating stroboscopic measurements as continuous measurements, the stochastic master equation (SME) describing the conditional evolution of the spin-oscillator state is
\begin{equation}\label{eq:1}
\begin{aligned}
d \rho =
(\mathcal{L}_{\mathrm{qba}}+&\mathcal{L}_{\mathrm{damp}})\rho\,dt
+ \kappa_p \sqrt{\eta}\,\mathcal{H}[\hat{p}_A]\rho\,dW_p(t) \\
&+ \kappa_x \sqrt{\eta}\,\mathcal{H}[\hat{x}_A]\rho\,dW_x(t),
\end{aligned}
\end{equation}
where $\mathcal{L}_{\mathrm{qba}}$ and $\mathcal{L}_{\mathrm{damp}}$ are the Lindblad terms
describing quantum back-action and interaction with the environment, respectively. The last two stochastic terms describe the state evolution of the spin oscillator conditioned on the measurement records $\mathbf{R}$, where $\mathcal{H}$ is the measurement superoperator, $W_x(t)$ and $W_p(t)$ are independent Wiener processes, and $\kappa_r^2$ is the measurement strength for observable $r$.

Eq.\eqref{eq:1} shows how the density matrix $\rho(t)$ represents our knowledge about the system by evolving through a sequence of changes accompanying each noisy  measurement. The subsequent measurements after time $t$ keep updating the density matrix as described by Eq.\eqref{eq:1}, but if we want to infer the outcome of a particular measurement {\it at} time $t$, its expected outcome is both governed by the density matrix at that time $\rho(t)$ and by our knowledge of later measurements as incorporated in the matrix $E(t)$. As shown in \cite{gammelmark2013past}, $E(t)$ is formally calculated by assuming that it equals the identity matrix $\mathbbm{1}$ beyond the final time of detection and by evolving it backward under the dynamics adjoint to Eq.\eqref{eq:1}, 
\begin{equation}\label{eq:3}
\begin{aligned}
    d E = (\mathcal{L}_{\mathrm{qba}}^\dagger+&\mathcal{L}^\dagger_{\mathrm{damp}})E d t + \kappa_p \sqrt{\eta} \mathcal{H} [\hat{p}_A^\dagger] E dW_p(t)+ \\ &\kappa_x \sqrt{\eta} \mathcal{H} [\hat{x}_A^\dagger] E dW_x(t),
\end{aligned}
\end{equation}

Rather than representing $\rho(t)$ and $E(t)$ as matrices in a discrete basis, we may employ the Gaussian character of the states and operations to fully characterize the states by the quadrature mean values and covariance matrix elements. Even though $E(t)$ is not a state, it is a positive definite matrix which permits a Gaussian Wigner function representation, and derives the corresponding retrodicted mean values and variances from the corresponding characterizations of $\rho(t)$ and $E(t)$ \cite{zhang2017prediction}.  

We thus define the first moments
$\langle r\rangle\equiv \operatorname{Tr}(r\rho)$ and the variances
$\sigma_r^2\equiv \langle r^2\rangle-\langle r\rangle^2$. Hereafter, $r\in\{x,p\}$ and $\bar{r}$ denotes the conjugate quadrature, with $\bar{x}=p$ and $\bar{p}=x$. The evolutions of the first moments and variances for $\hat{x}_{\mathrm A}$ and $\hat{p}_{\mathrm A}$ are
\begin{equation}\label{eq:first_moments}
\begin{aligned}
d\langle r\rangle
=-\Gamma_{\mathrm{tot}}\langle r\rangle\,dt
+\sqrt{2\eta}\,\kappa_r\sigma_r^2\,dW_r(t),
\end{aligned}
\end{equation}
and
\begin{equation}\label{eq:variances}
\begin{aligned}
\frac{d\sigma_r^2}{dt} = -2\Gamma_{\mathrm{tot}}\sigma_r^2 + 2\Gamma_{\mathrm{tot}}V_0 + \frac{\kappa_{\bar{r}}^2}{2} - 2\eta\kappa_r^2\sigma_r^4,
\end{aligned}
\end{equation}
where $\Gamma_{\mathrm{tot}}$ denotes the total damping rate of the system and $V_0$ is the bath variance due to imperfect optical pumping and environmental noise. The first two terms describe dissipation and noise of the state with imperfect pumping, the third represents quantum back-action, and the last term captures the noise reduction arising from information gained through measurements. The prediction-only trajectory shown in Fig.~\ref{Fig:1}d (left panel) is reconstructed
from an individual measurement record. For a given record, the measurement outcomes condition the state \(\rho(t)\), giving the conditional estimates \((\langle \hat{x}_{\mathrm A}(t)\rangle,
\langle \hat{p}_{\mathrm A}(t)\rangle ) \) plotted in phase space. Notably, the trajectory is a single-shot filtered reconstruction of the oscillator motion, not an ensemble average over repeated experiments and should also not be interpreted as the outcome of a joint projective measurement of \(\hat{x}_{\mathrm A}\) and \(\hat{p}_{\mathrm A}\). The corresponding prediction errors, shown in Fig.~\ref{Fig:2}a, are obtained
by comparing the filtered estimates with independent weak verification
measurements of the same quadratures. 

The steady-state solutions of the variances for $\hat{x}_{\mathrm{A}}$ and $\hat{p}_{\mathrm{A}}$ in prediction are,
\begin{equation}\label{eq:2}
\sigma_r^2
=\frac{\sqrt{1+4V_0\kappa_r^2\eta/\Gamma_{\mathrm{tot}}+\kappa_r^2\kappa_{\bar{r}}^2\eta/\Gamma_{\mathrm{tot}}^2}-1}{2\kappa_r^2\eta/\Gamma_{\mathrm{tot}}}.
\end{equation}
The steady-state prediction variance in Fig.~\ref{Fig:2}a is slightly above the SQL mainly due to imperfections of spin polarization. Under balanced measurement strengths ($\kappa_x = \kappa_p$), perfect pumping ($V_0 =0.5$) and perfect quantum detection efficiency $\eta = 1$, the steady-state variance reduces to $\sigma^2_{x_{\mathrm{A}}/p_{\mathrm{A}}} = 0.5$, corresponding to the variance of the coherent spin state (CSS) of the spin oscillator, and reflecting the symmetry of the balanced weak measurements applied to the two quadratures. We emphasize that this is fundamentally different from QND quadrature measurements where single-sided (i.e., prediction only) estimation implies squeezing of the measured quadrature below CSS. Our alternating measurements of $x_{\mathrm{A}}$ and $p_{\mathrm{A}}$ entail an effective ``squeezing-unsqueezing" effect (as shown in ~\ref{Fig:1}c) preventing the variances from going below CSS.

\begin{figure}[t] 
	\centering
	\includegraphics[width=3.5in]{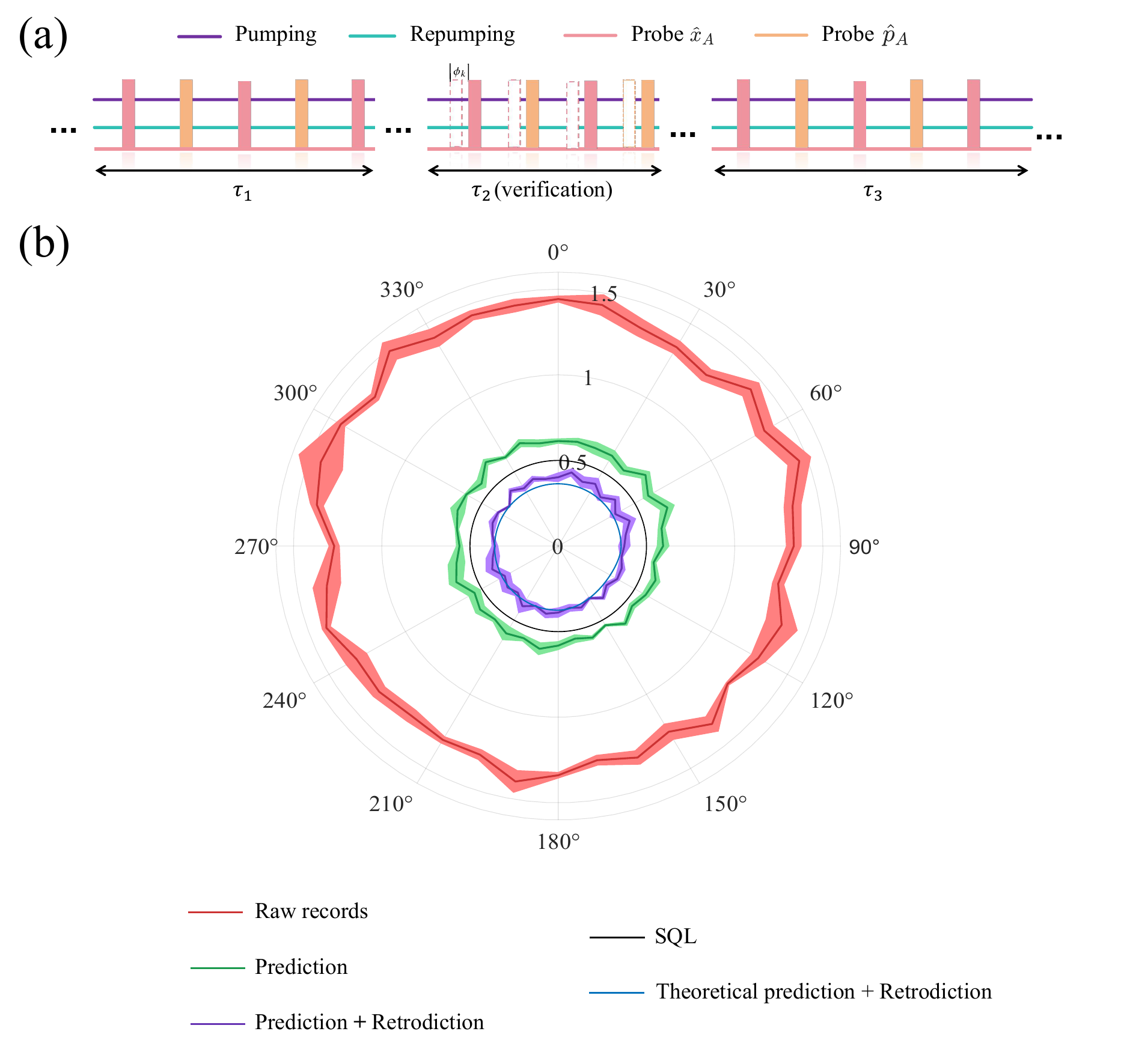}	
	\caption{{\bf Spin-noise variance below the standard quantum limit (SQL) in all directions.}
		{\bf (a) Pulse sequence for angle-resolved measurement.} { The pump beam remains continuously on to maintain steady-state conditions. The probe beam is divided into three temporal segments: prediction period $\tau_1$, verification period $\tau_2$ (used to confirm squeezing), and retrodiction period $\tau_3$. The phase of the stroboscopic probe light is aligned during $\tau_1$ and $\tau_3$, while $\tau_2$ is phase-delayed by an angle $\theta$.}
		{\bf (b) Angular dependence of spin-noise variance.}{ Spin noise variance as a function of quadrature angle for the cases of raw measurement data (red), prediction-based squeezing (green), and past quantum state squeezing (purple). Shaded error bands denote one standard deviation, derived from five independent experimental runs, each comprising $1\times10^{4}$ repetitions of the pulse sequence. The variance of the SQL is 0.5 (black circle); theoretically, the variance when both prediction and retrodiction are considered is 0.37 (blue circle).}
		}
	\label{Fig:3}			
\end{figure}

Similar equations apply to first and second moments of the Gaussian ``state'' $E$, leading to the backward evolution shown in Fig.\:\ref{Fig:2}a during which the variances reach steady-state values,
\begin{equation}\label{eq:6}
\gamma_r^2 =
\frac{\sqrt{1+4V_0\kappa_r^2\eta/\Gamma_{\mathrm{tot}}+\kappa_r^2\kappa_{\bar{r}}^2\eta/\Gamma_{\mathrm{tot}}^2}+1}
{2\kappa_r^2\eta/\Gamma_{\mathrm{tot}}}.
\end{equation}
It can be derived that~\cite{bao2020retrodiction,zhang2017prediction,khanahmadi2021guessing} the retrodicted variance $\Delta_r^2$ obeys $1/\Delta_r^{2} = 1/\sigma_r^{2} + 1/\gamma_r^{2}$, which yields, 
\begin{equation}\label{eq:5}
\Delta_r^2 =
\frac{4V_0+\kappa_{\bar{r}}^2/\Gamma_{\mathrm{tot}}}
{4\sqrt{1+4V_0\kappa_r^2\eta/\Gamma_{\mathrm{tot}}+\kappa_r^2\kappa_{\bar{r}}^2\eta/\Gamma_{\mathrm{tot}}^2}}.
\end{equation}
The dependence of $\Delta_{x_{\mathrm{A}}/p_{\mathrm{A}}}^{2}$ on $\kappa^2/\Gamma_{tot}$ is plotted in Fig.\:\ref{Fig:2}b. The results show that our method can ideally achieve a maximum squeezing of $-3$\:dB on the $\hat{x}_{\mathrm{A}}/\hat{p}_{\mathrm{A}}$ simultaneously, or go beyond the Heisenberg uncertainty limit by 3 dB in the sense of conditional estimation. 

Using the experimental configuration shown in Fig.~\ref{Fig:1}b, we acquire continuous optical measurement records of the spin-oscillator state over 1.4 hours. From these records, the phase-space trajectories are reconstructed and Fig~\ref{Fig:1}d shows a representative 31.8-ms segment of the reconstructed trajectories. The right panel shows the trajectory obtained by combining prediction and retrodiction. In comparison with the left panel, it exhibits smaller fluctuations due to the inclusion of information from the future measurement record. 

Fig.\:\ref{Fig:2}c compares the Wineland squeezing parameters $\xi_W^2$ for $\hat{x}_{\mathrm{A}}$ and $\hat{p}_{\mathrm{A}}$ derived from: (i) raw measurement data, (ii) the deviation from predictions by  $\rho(t)$, and (iii) the deviation from the values retrodicted by $\rho(t)$ and $E(t)$. The raw recorded data is noisy, as expected because of the back-action noise originated from the interaction. The variance of the conditional quantum state $\rho(t)$ is ideally at the SQL but is in practice slightly higher due to experimental imperfections. By retrodiction we can state values of both non-commuting observables $\hat{x}_{\mathrm{A}}$ and $\hat{p}_{\mathrm{A}}$ with sub-SQL uncertainties, and we measure $\Delta x \cdot \Delta p$ is $-1.01 \pm 0.15\:{\rm dB}$, lower than the HUR for over 1.4 hours. Notably, what is inferred with high precision is not joint projective measurement outcomes, but the conditional estimates of both non-commuting observables reconstructed from continuous weak measurement records. The corresponding variances are obtained from the statistics of weak measurements. In contrast, if one applies a projective measurement to one component, the squeezing of the conjugate component will be destroyed by quantum back-action. The weak measurements used to verify these variances will also induce small quantum back-action for including a few alternating $x$ and $p$ measurement pulses, which is analyzed in detail in the Supplementary Information. Similar results have been observed using a quantum memory on several bits in an optical system and by similar retrodiction as ours in a spin system \cite{bao2020retrodiction,rossi2019observing,li2011experimental,prevedel2011experimental}. In contrast, our experiment achieves \textit{concurrent} sub-SQL variances with direct access to both observables through continuous weak measurement records.

The past quantum state formalism further enables Wigner function reconstruction for arbitrary quadratures of the spin oscillator \cite{SI}. By incrementally varying the stroboscopic probe phase (Fig.\:\ref{Fig:3}a) across $1\times10^4$ experimental realizations, we measure spin noise distributions throughout the $J_y$--$J_z$ plane. The measured atomic spin noise is shown in Fig.\:\ref{Fig:3}b, and is scaled to atomic projection noise. The raw recorded data and the quantum state $\rho$ are noisier than the SQL in all directions. Using the past quantum state, the variance of the spin oscillator could go below the SQL in arbitrary directions, and the average noise of the quadratures is $0.388\pm 0.025$ compared to the SQL of 0.5. To our knowledge, this is the first experimental demonstration of sub-SQL conditional estimation over all phase-space quadratures. For comparison, in our previous work the information of $x$ and $p$ is obtained before and after $t$ respectively, resulting in a butterfly-like distribution where only the noise along the $x$ and $p$ axes was confirmed to be below the SQL \cite{bao2020retrodiction}.

\begin{figure}[t]
	\centering
	\includegraphics[width=3.5in]{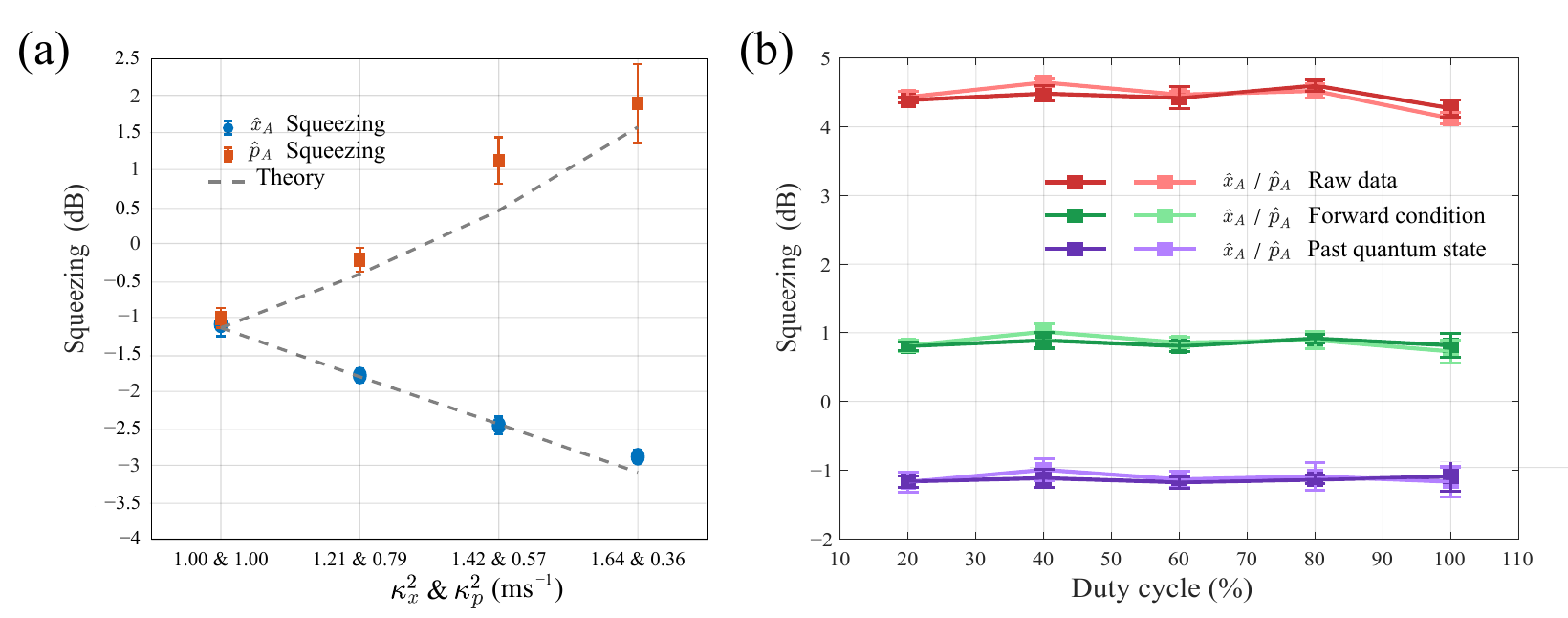}	
	\caption{{\bf Degree of squeezing versus probe intensity and duty cycle.}
    {\bf (a) Dependence of squeezing on the measurement strength. }	When $\kappa_x^2=\kappa_p^2$, the $\hat{x}_A$ and $\hat{p}_A$ quadratures exhibit the same degree of squeezing. Keeping the total optical power constant, as $\kappa_x^2$ decreases and $\kappa_p^2$ increases, the squeezing degree in the $\hat{x}_A$ quadrature gradually diminishes until anti-squeezing is observed, while the squeezing in the $\hat{p}_A$ quadrature continuously increases. The larger error bar on the last point of $\hat{x}_A$ is due to the limited measurement strength.
    {\bf (b) Dependence of the squeezing on the duty cycle.}{ The degrees of squeezing of both $\hat{x}_A$ and $\hat{p}_A$ are insensitive to the temporal duty cycle.}}
	\label{Fig:4}			
\end{figure}

We now study the squeezing versus the measurement strength of $\hat{x}_{\mathrm{A}}$ and $\hat{p}_{\mathrm{A}}$. We keep the average probe power unchanged and vary the ratio of the probe powers measuring $\hat{x}_{\mathrm{A}}$ and $\hat{p}_{\mathrm{A}}$. The results of squeezing are shown in Fig.\:\ref{Fig:4}a. When the measurement strength ratio deviates from unity, the squeezing becomes asymmetric: the increase of measurement strength on $\hat{x}_{\mathrm{A}}$ will bring more quantum back-action noise on $\hat{p}_{\mathrm{A}}$ and strengthen the noise reduction effect of prediction and retrodiction, and vice versa. When the ratio is $1.64/0.36$, the $\hat{x}_{\mathrm{A}}$ is $-2.88 \pm 0.087$ \:dB squeezing and $\hat{p}_{\mathrm{A}}$ is $1.89\pm 0.53$ \:dB anti-squeezing. This trade-off explicitly demonstrates the measurement-disturbance relationship inherent in non-commuting observables, even when employing retrodiction-enhanced estimation. 

To extend our protocol from the quasi-continuous regime to the truly continuous case, we increased the duty cycle of probes up to the continuous-measurement limit, and kept the measurement strength fixed so that the effective measurement rate remained constant. In this scenario, the continuously measured spin component corresponds to a quadrature that rotates through phase space due to the Larmor precession. As shown in Fig.\:\ref{Fig:4}b, for fixed $\kappa^2\eta$ and $\Gamma_{tot}$, both the unconditioned and conditionally inferred steady-state variances are insensitive to the temporal duty cycle, which aligns well with our calculations \cite{SI}. These results demonstrate that our method does not depend on the stroboscopic technique and could naturally extend to continuous readout or other measurement schemes without sacrificing squeezing performance.

In conclusion, we have experimentally demonstrated, for the first time, the simultaneous sub-SQL estimation of two non-commuting observables in a steady-state spin oscillator using quantum trajectory theory. By continuously monitoring both canonical quadratures through quantum non-demolition measurements and combining prediction with retrodiction within the past quantum state formalism, we achieve conditional variances for $\hat{x}$ and $\hat{p}$—as well as for arbitrary quadratures in phase space—that lie below the standard quantum limit. The central ingredient of this approach is the consistent incorporation of both past and future measurement information to refine the inference of the system state at a given time, while properly accounting for measurement-induced back-action. Our results establish quantum trajectory theory as a powerful framework for overcoming conventional measurement bounds in continuous monitoring scenarios and provide an operational route toward multi-parameter quantum metrology beyond the standard quantum limit \cite{cao2025joint,li2026multiparameter,zhang2026distributed}. The protocol could achieve higher squeezing in the quadratures approaching the 3 dB theoretical limit, with an optical cavity \cite{ma2022high}, improved optical efficiency, or an adaptive measurement strategy \cite{kaubruegger2019variational}. Furthermore, entangling with an ancillary atomic ensemble can allow more than 3 dB enhancement in arbitrary quadratures of the $x$-$p$ phase space \cite{khanahmadi2021guessing}.

\begin{acknowledgments}
We thank Howard Wiseman and Andrew Jordan for helpful discussions. This work is supported by National Natural Science Foundation of China under Grant No. 12027806 and No. 12161141018. K. M. acknowledges support from the Carlsberg Foundation through the Semper Ardens project QcooL.
\end{acknowledgments}

\bibliography{scibib}

\end{document}


\title{Supplementary Information for \\
Quantum trajectory of a spin oscillator, continuous retrodiction beyond the Heisenberg uncertainty relation}

\author{Peiying Li}%
\affiliation{State Key Laboratory of Quantum Optics Technologies and Devices, Institute of Laser Spectroscopy, Shanxi University, Taiyuan, Shanxi 030006, China}%
\affiliation{Collaborative Innovation Center of Extreme Optics, Shanxi University, Taiyuan 030006, China}
\author{Junlei Duan}%
\email{20110190074@fudan.edu.cn}
\affiliation{State Key Laboratory of Surface Physics, Department of Physics and Key Laboratory of Micro and Nano Photonic Structures (Ministry of Education), Fudan University, Shanghai 200433, China}%
\author{Suotang Jia}%
\affiliation{State Key Laboratory of Quantum Optics Technologies and Devices, Institute of Laser Spectroscopy, Shanxi University, Taiyuan, Shanxi 030006, China}%
\affiliation{Collaborative Innovation Center of Extreme Optics, Shanxi University, Taiyuan 030006, China}
\author{Klaus M\o lmer}%
\email{klaus.molmer@nbi.ku.dk} 
\affiliation{Niels Bohr Institute, University of Copenhagen, Jagtvej 155A, DK 2200 Copenhagen, Denmark}%
\author{Yanhong Xiao}%
\email{yxiao@fudan.edu.cn}
\affiliation{State Key Laboratory of Surface Physics, Department of Physics and Key Laboratory of Micro and Nano Photonic Structures (Ministry of Education), Fudan University, Shanghai 200433, China}%
\affiliation{State Key Laboratory of Quantum Optics Technologies and Devices, Institute of Laser Spectroscopy, Shanxi University, Taiyuan, Shanxi 030006, China}%
\affiliation{Collaborative Innovation Center of Extreme Optics, Shanxi University, Taiyuan 030006, China}%
	
\maketitle

\tableofcontents

\newpage

\section*{S1\quad Atom-light interaction}

In atom-light interactions involving the polarization states of light, the quantum state of light is conventionally characterized by Stokes operators $\hat{S}_x$, $\hat{S}_y$ and $\hat{S}_z$. Meanwhile, the quantum state of atomic spins is represented through collective atomic operators $\hat{J}_i=\sum_k \hat{j}^k_i$ $(i=x,y,z)$, defined as the ensemble sum of individual spin operators, enabling the description of collective spin systems. For a monochromatic light field propagating along the $z$-axis, its polarization properties are described by the Stokes operators, which are defined in terms of the annihilation ($\hat{a}$) and creation ($\hat{a}^\dagger$) operators for the two orthogonal polarization modes (denoted $x$ and $y$):
\begin{equation}\label{eq:1}
\begin{aligned}
\hat{S}_x &= \frac{1}{2} \left( \hat{a}_x^\dagger \hat{a}_x - \hat{a}_y^\dagger \hat{a}_y \right) = \frac{1}{2} \left( \hat{n}_x - \hat{n}_y \right), \\
\hat{S}_y &= \frac{1}{2} \left( \hat{a}_x^\dagger \hat{a}_y + \hat{a}_y^\dagger \hat{a}_x \right) = \frac{1}{2} \left( \hat{n}_{+45^\circ} - \hat{n}_{-45^\circ} \right), \\
\hat{S}_z &= \frac{1}{2i} \left( \hat{a}_x^\dagger \hat{a}_y - \hat{a}_y^\dagger \hat{a}_x \right) = \frac{1}{2} \left( \hat{n}_{\sigma_+} - \hat{n}_{\sigma_-} \right),
\end{aligned}
\end{equation}
where $\hat{n}_{polarization}$ is the photon number operator, and the polarization state can be described using linear polarization states basis along the $x$ and $y$ axes, or ±45° diagonal axes, as well as left- and right-handed circular polarization in the laboratory coordinate system.

In the experiment, continuous optical pumping is collinear with the bias magnetic field, the atomic ensemble is polarized into an end state of the Zeeman sublevels. For such a highly polarized atomic ensemble, the Holstein-Primakoff transformation can be employed to map atomic operators and light operators \cite{HP1940}. The canonical operators for the optical field can be expressed as:
\begin{equation}\label{eq:2}
\begin{aligned}
\hat{x}_{L} = \frac{\hat{S}_{y}}{\sqrt{|\langle S_{x}\rangle|}}, 
\quad 
\hat{p}_{L} = \frac{\hat{S}_{z}}{\sqrt{|\langle S_{x}\rangle|}},
\end{aligned}
\end{equation}
and the canonical operators for the collective atomic spin polarized along the $x$-direction can be written as:
\begin{equation}\label{eq:3}
\begin{aligned}
\hat{x}_{A} = \frac{\hat{J}_{y}}{\sqrt{|\langle J_{x}\rangle|}}, 
\quad 
\hat{p}_{A} = \frac{\hat{J}_{z}}{\sqrt{|\langle J_{x}\rangle|}},
\end{aligned}
\end{equation}
obeying $[\hat{x}_{L}, \hat{p}_{L}] = i$ and $[\hat{x}_{A}, \hat{p}_{A}] = i$, where we set $\hbar = 1$.

In our experiment, the probe beam is coupled to the  $^{87}$Rb $D_2$ transition with a blue detuning of 4.5 GHz, realizing a standard implementation of quantum non-demolition (QND) measurement~\cite{julsgaard2003entanglement}. The QND Hamiltonian is:
\begin{equation}\label{eq:4}
\begin{aligned}
\hat{H}_{QND}= \kappa \hat{p}_A\hat{p}_L,
\end{aligned}
\end{equation}
where $\kappa$ represents the coupling strength between light and atoms.

\section*{S2\quad Quantum trajectory theory} 
\label{sec:s2}
We first analyze the evolution of the density matrix for the Gaussian state of the spin oscillator in this section. Optical pumping, spin relaxation, and residual technical noise are modeled as coupling to an effective thermal reservoir with steady-state variance $V_0$, while the probe measurements provide both information gain and quantum back-action. In the rotating frame, the stochastic master equation for the conditioned density matrix $\rho$ can be written as
\begin{equation}\label{eq:s2_sme}
    d \rho = (\mathcal{L}_{qba}+ \mathcal{L}_{damp})\rho d t + \kappa_p \sqrt{\eta} \mathcal{H} [\hat{p}_A] \rho dW_p(t) + \kappa_x \sqrt{\eta} \mathcal{H} [\hat{x}_A] \rho dW_x(t),
\end{equation}
where $\kappa_x^2$ and $\kappa_p^2$ are the measurement strengths of $\hat{x}_{\mathrm{A}}$ and $\hat{p}_{\mathrm{A}}$, respectively, $\eta$ is the total detection efficiency, and $dW_x$ and $dW_p$ are two independent Wiener increments satisfying $dW_i dW_j = \delta_{ij}dt$. We use the standard Lindblad dissipator and measurement superoperators
\begin{equation}
\mathcal{D}[\hat{o}]\rho
=
\hat{o}\rho \hat{o}^{\dagger}
-\frac{1}{2}\hat{o}^{\dagger}\hat{o}\rho
-\frac{1}{2}\rho\hat{o}^{\dagger}\hat{o},
\end{equation}
and
\begin{equation}
\mathcal{H}[\hat{o}]\rho = \frac{1}{\sqrt{2}}\{ \hat{o}\rho + \rho\hat{o}^{\dagger} - \mathrm{Tr}\!\left[(\hat{o}+\hat{o}^{\dagger})\rho\right]\rho\}.
\end{equation}
The Liouvillian terms \(\mathcal{L}_{\mathrm{qba}}\) and 
\(\mathcal{L}_{\mathrm{damp}}\) generally contain several Lindblad channels. 
For example, the probe-induced quantum back-action can be written as
\begin{equation}
\mathcal{L}_{\mathrm{qba}}\rho
=
\frac{\kappa_x^2}{2}\mathcal{D}[\hat{x}_{\mathrm A}]\rho
+
\frac{\kappa_p^2}{2}\mathcal{D}[\hat{p}_{\mathrm A}]\rho.
\end{equation}
For the Gaussian state relevant to our experiment, the conditional dynamics of the first moments are
\begin{equation}\label{eq:s2_first_moments}
\begin{aligned}
d\langle x\rangle &= -\Gamma_{\mathrm{tot}}\langle x\rangle\,dt
+ \sqrt{2\eta}\kappa_x \sigma_x^2\, dW_x,\\
d\langle p\rangle &= -\Gamma_{\mathrm{tot}}\langle p\rangle\,dt
+ \sqrt{2\eta}\kappa_p \sigma_p^2\, dW_p.
\end{aligned}
\end{equation}
While the covariance matrix remains diagonal, the variances of the two quadratures obey closed Riccati equations. Using It\^o calculus, we obtain
\begin{equation}\label{eq:s2_varx}
\frac{d\sigma_x^2}{dt}= -2\Gamma_{\mathrm{tot}}\sigma_x^2 + 2\Gamma_{\mathrm{tot}}V_0 + \frac{\kappa_p^2}{2} - 2\kappa_x^2\eta \sigma_x^4 ,
\end{equation}
\begin{equation}\label{eq:s2_varp}
\frac{d\sigma_p^2}{dt}= -2\Gamma_{\mathrm{tot}}\sigma_p^2 + 2\Gamma_{\mathrm{tot}}V_0 + \frac{\kappa_x^2}{2} - 2\kappa_p^2\eta \sigma_p^4 .
\end{equation}
The first two terms describe damping toward the unconditional steady-state variance $V_0$, the third term is the probe back-action, and the last term is the information gain from the measurement record. Solving Eqs.~(\ref{eq:s2_varx}) and (\ref{eq:s2_varp}) in the steady state gives
\begin{equation}\label{eq:s2_ssx}
\sigma_x^2 = \frac{\sqrt{1+4V_0\kappa_x^2\eta/\Gamma_{\mathrm{tot}}+\kappa_x^2\kappa_p^2\eta/\Gamma_{\mathrm{tot}}^2}-1}{2\kappa_x^2\eta/\Gamma_{\mathrm{tot}}},
\end{equation}
\begin{equation}\label{eq:s2_ssp}
\sigma_p^2 = \frac{\sqrt{1+4V_0\kappa_p^2\eta/\Gamma_{\mathrm{tot}}+\kappa_x^2\kappa_p^2\eta/\Gamma_{\mathrm{tot}}^2}-1}{2\kappa_p^2\eta/\Gamma_{\mathrm{tot}}}.
\end{equation}
These expressions reproduce the prediction variances used in the main text. In particular, when $\kappa_x=\kappa_p$, the two quadratures acquire the same conditional variance; when the two strengths are unbalanced, one quadrature is squeezed more strongly at the price of increased noise in the other one.

Quantum trajectory theory also incorporates information carried by future measurements through the effect matrix $E$ \cite{gammelmark2013past,zhang2017prediction}. In the same rotating-frame description, $E$ obeys the adjoint stochastic master equation,
\begin{equation}\label{eq:s2_effect}
\begin{aligned}
    d E = (\mathcal{L}_{qba}^\dagger+ \mathcal{L}^\dagger_{damp})E d t + \kappa_p \sqrt{\eta} \mathcal{H} [\hat{p}_A^\dagger] E dW_p(t)+  \kappa_x \sqrt{\eta} \mathcal{H} [\hat{x}_A^\dagger] E dW_x(t),
\end{aligned}
\end{equation}
with $E(T)=\mathbbm{1}$ at a final time $T$. Here
\begin{equation}
\mathcal{D}^{\dagger}[\hat{o}]E = \hat{o}^{\dagger}E\hat{o} - \frac{1}{2}\hat{o}^{\dagger}\hat{o}E - \frac{1}{2}E\hat{o}^{\dagger}\hat{o}.
\end{equation}
The $E$ operator can be assigned a Gaussian Wigner function, and the reversed dynamics of $E$ is thereby characterized by  the first moment equations,
\begin{equation}\label{eq:s2_retro_first_moments}
\begin{aligned}
d\langle x\rangle &= \Gamma_{\mathrm{tot}}\langle x\rangle\,dt
+ \sqrt{2\eta}\kappa_x \gamma_x^2\, dW_x,\\
d\langle p\rangle &= \Gamma_{\mathrm{tot}}\langle p\rangle\,dt
+ \sqrt{2\eta}\kappa_p \gamma_p^2\, dW_p.
\end{aligned}
\end{equation}
The Wigner function of the operator $E$  is further characterized by variances $\gamma_x^2$ and $\gamma_p^2$ satisfying the backward equations,
\begin{equation}\label{eq:s2_retx}
\frac{d\gamma_x^2}{dt}= 2\Gamma_{\mathrm{tot}}\gamma_x^2 + 2\Gamma_{\mathrm{tot}}V_0 + \frac{\kappa_p^2}{2} - 2\kappa_x^2\eta \gamma_x^4,
\end{equation}
\begin{equation}\label{eq:s2_retp}
\frac{d\gamma_p^2}{dt}= 2\Gamma_{\mathrm{tot}}\gamma_p^2 + 2\Gamma_{\mathrm{tot}}V_0 + \frac{\kappa_x^2}{2} - 2\kappa_p^2\eta \gamma_p^4,
\end{equation}
which lead to the steady-state solutions,
\begin{equation}\label{eq:s2_retssx}
\gamma_x^2 = \frac{\sqrt{1+4V_0\kappa_x^2\eta/\Gamma_{\mathrm{tot}}+\kappa_x^2\kappa_p^2\eta/\Gamma_{\mathrm{tot}}^2}+1}{2\kappa_x^2\eta/\Gamma_{\mathrm{tot}}},
\end{equation}
\begin{equation}\label{eq:s2_retssp}
\gamma_p^2 = \frac{\sqrt{1+4V_0\kappa_p^2\eta/\Gamma_{\mathrm{tot}}+\kappa_x^2\kappa_p^2\eta/\Gamma_{\mathrm{tot}}^2}+1}{2\kappa_p^2\eta/\Gamma_{\mathrm{tot}}}.
\end{equation}
Combining the information accumulated in $\rho$ and $E$  yields the past quantum state, for which the conditional variance of either quadrature is obtained from the inverse-variance sum \cite{zhang2017prediction},
\begin{equation}\label{eq:s2_pqsvar}
\left(\Delta_{x_{\mathrm{A}}/p_{\mathrm{A}}}^{2}\right)^{-1}=\left(\sigma_{x_{\mathrm{A}}/p_{\mathrm{A}}}^{2}\right)^{-1}+\left(\gamma_{x_{\mathrm{A}}/p_{\mathrm{A}}}^{2}\right)^{-1}.
\end{equation}
The above results describe the conditional and retrodictive uncertainties associated with continuous measurements. To further generalize the description of the measurement process, in the next section we adopt the positive operator-valued measure (POVM) in the past quantum state formalism. This enables us to analyze the estimation of $\hat{x}_A$ and $\hat{p}_A$, as well as quadratures along arbitrary directions in a weak measurement regime, and also to quantify the corresponding estimation errors in a unified framework.

\section*{S3\quad Past quantum state}

Considering the large number of atoms and photons in the experiment, the collective atomic spin and the optical field can be linearized around their mean values. The corresponding canonical operators thus describe quantum fluctuations that can be approximated as Gaussian states with zero mean and variance $1/2$:
\begin{equation}\label{eq:s3_initial_wavefunction}
\begin{aligned}
\psi_{\hat{p}_{\mathrm{A}}}(x) = \psi_{\hat{x}_{\mathrm{L}}}(x) = \psi(x) = \frac{1}{\pi^{1/4}} \exp\left(-\frac{x^2}{2}\right).
\end{aligned}
\end{equation}
We define the eigenstates of the atomic quadrature operator $\hat{p}_A$ and the optical quadrature operator $\hat{x}_L$ as
\begin{equation}
\begin{aligned}
\hat{p}_A |a\rangle_{\hat{p}_A} &= a |a\rangle_{\hat{p}_A}, \\
\hat{x}_L |l\rangle_{\hat{x}_L} &= l |l\rangle_{\hat{x}_L},
\end{aligned}
\end{equation}
where $a$ and $l$ denote the corresponding eigenvalues.

The initial atomic and optical states, $\Psi_{\mathrm{A}}^{\mathrm{in}}$ and $\Psi_{\mathrm{L}}^{\mathrm{in}}$, can be expanded in these quadrature eigenbases as
\begin{equation}
\begin{aligned}
\Psi_{\text{A}}^{\text{in}} &= \int \mathrm{d}a \, \psi_{\hat{p}_{\text{A}}}(a) |a\rangle_{\hat{p}_{\text{A}}}, \\
\Psi_{\text{L}}^{\text{in}} &= \int \mathrm{d}l \, \psi_{\hat{x}_{\text{L}}}(l) |l\rangle_{\hat{x}_{\text{L}}},
\end{aligned}
\end{equation}
with $\psi_{\hat{p}_A}(a)$ and $\psi_{\hat{x}_L}(l)$ denoting the corresponding wavefunctions in the $\hat{p}_A$ and $\hat{x}_L$ representations, respectively.

POVM (positive-operator-valued measure) is used to describe the interaction between light field and atomic system~\cite{wiseman2010quantum}, the initial atomic state is described by a density matrix \(\rho_0\). The initial atomic state is taken as $\Psi_{A}^{in}$, then the corresponding density matrix is given by $\rho_0 = |\Psi_A^{in}\rangle\langle\Psi_A^{in}|$. For QND interaction $\hat{C}_z = e^{-i \widetilde{\kappa}\hat{p}_A\otimes\hat{p}_L}$, where $\widetilde{\kappa}$ is the dimensionless effective coupling strength characterizing the total QND interaction accumulated during the measurement interval. The measurement operator is defined as \cite{bao2020retrodiction}:
\begin{equation}\label{eq:s3_povm_operator}
\begin{aligned}
  \hat{\Omega}_m = (\hat{I}_A \otimes \langle m|_{\hat{x}_L}) \hat{C}_z (\hat{I}_A \otimes |\Psi_L^{in}\rangle).
\end{aligned}
\end{equation}
For an arbitrary spin direction $\hat{x}_A(\theta) = \hat{p}_A\cos\theta + \hat{x}_A\sin\theta$,
\begin{equation}\label{eq:s3_povm_operator_theta}
\begin{aligned}
\hat{\Omega}_m = \int \psi(m - \widetilde{\kappa} a) |a,\theta\rangle\langle a,\theta| da,
\end{aligned}
\end{equation}
where $|a,\theta\rangle$ is the eigenstate of $\hat{x}_A(\theta)$.

We divide the continuous measurement record into three sections, represented by the effective measurement outcomes $m_1$, $m_2$, and $m_3$: the heralding record, the verification record, and the postselection record, respectively. The heralding outcome $m_1$ conditions the forward-evolving state $\rho$ and thereby enables prediction of the verification outcome $m_2$, whereas the postselection outcome $m_3$ provides the future measurement information encoded in the backward-evolving effect matrix $E$, thereby enabling retrodiction of the verification outcome $m_2$ \cite{gammelmark2013past}. In forward prediction, given the $m_1$, the conditional probability for the verification measurement is:
\begin{equation}
\begin{aligned}
\mathrm{Pr}(m_2|m_1)
&=
\frac{\mathrm{Tr}\left(
\hat{\Omega}_{m_2}\rho\hat{\Omega}_{m_2}^{\dagger}
\right)}
{\int \mathrm{d}m_2'\,
\mathrm{Tr}\left(
\hat{\Omega}_{m_2'}\rho\hat{\Omega}_{m_2'}^{\dagger}
\right)}
\\
&\propto
\int \mathrm{d}a\,
\left|\psi(m_2-\widetilde{\kappa}_2 a)\right|^2
\langle a,\theta|\rho|a,\theta\rangle ,
\end{aligned}
\end{equation}
where $\rho = \hat{\Omega}_{m_1}\rho_0\hat{\Omega}_{m_1}^\dagger$.

If both prediction and retrodiction are considered, then conditioning on the measurement results $m_1$ and $m_3$, the conditional probability for the second measurement result is given by:
\begin{equation}
\begin{aligned}
\mathrm{Pr}(m_2|m_1,m_3)
&=
\frac{
\operatorname{Tr}\left(
\hat{\Omega}_{m_2}\rho
\hat{\Omega}_{m_2}^{\dagger}E
\right)
}{
\int \mathrm{d}m_2'\,
\operatorname{Tr}\left(
\hat{\Omega}_{m_2'}\rho
\hat{\Omega}_{m_2'}^{\dagger}E
\right)
}
\\
&\propto
\iint \mathrm{d}a\,\mathrm{d}a'\,
\psi(m_2-\widetilde{\kappa}_2 a)
\psi^{*}(m_2-\widetilde{\kappa}_2 a')
\\
&\qquad\times
\langle a,\theta|\rho|a',\theta\rangle
\langle a',\theta|E|a,\theta\rangle .
\end{aligned}
\end{equation}

Within the framework of quantum trajectory theory, the state of the system is jointly described by the forward-evolved density matrix $\rho(t)$ and the effect matrix $E(t)$ that incorporates information from future measurements. 
Backward effect matrix $E$ is defined solely by the third measurement $m_3$ in the three-measurement sequence:
\begin{equation}
\begin{aligned}
E = \hat{\Omega}_{m_3}^\dagger \hat{\Omega}_{m_3}.
\end{aligned}
\end{equation}

The quasi-probability distributions of these two operators in phase space are represented by their Wigner functions.
The Wigner function \(W_\rho(q, p)\) of the forward density matrix \(\rho\) is defined as a Gaussian function, which encodes the constraints imposed by all measurements from the initial time up to the current time \(t\):
\begin{equation}\label{eq:Wigner_rho}
\begin{aligned}
W_{\rho}(q, p) = \frac{1}{\pi} \int \mathrm{~d} r \cdot\langle p+r, 0| \rho|p-r, 0\rangle e^{-2 i q r} \propto \frac{1}{2 \pi \sigma_1 \sigma_2} \exp\left( -\frac{p^2}{2\sigma_1^2} -\frac{q^2}{2\sigma_2^2}\right).
\end{aligned}
\end{equation}

Similarly, the Wigner function \(W_E(q, p)\) of the backward effect matrix \(E\) is also Gaussian, encoding the constraints from all future measurements from time \(t\) to the final time \(T\):
\begin{equation}\label{eq:Wigner_E}
\begin{aligned}
W_E(q, p)=\frac{1}{\pi} \int \mathrm{~d} r \cdot\langle q+r, \pi / 2| E|q-r, \pi / 2\rangle e^{2 i p r} \propto \frac{1}{2 \pi \sigma_3 \sigma_4} \exp\left( -\frac{p^2}{2\sigma_3^2} -\frac{q^2}{2\sigma_4^2}\right).
\end{aligned}
\end{equation}
Here, the canonical variables \(q\) and \(p\) correspond to the eigenvalues of the atomic spin quadratures \(\hat{x}_\mathrm{A}\) and \(\hat{p}_\mathrm{A}\), respectively. The parameters $\sigma_{1,2,3,4}$ denote the standard deviations of the quadratures associated with the Wigner functions $W_{\rho}$ and $W_E$. To derive the conditional variance along an arbitrary spin direction $\theta$, we need to compute the matrix elements of $\rho$ and $E$ in the rotated basis $|v,\theta\rangle$, where $|v,\theta\rangle$ is the eigenstate of the rotated spin operator $\hat{x}_A(\theta)=\hat{p}_A\cos\theta+\hat{x}_A\sin\theta$ with eigenvalue $v$. According to the Wigner-Weyl transform, these matrix elements can be obtained by integrating the Wigner functions with appropriate phase factors:
\begin{equation}
\langle v,\theta | \rho | v',\theta \rangle = \int_{-\infty}^{\infty} du \, W_\rho(q,p) \, e^{i u (v - v')},
\label{eq:matrix_element_rho_integral}
\end{equation}
where the phase-space coordinates $ (q,p)$ are expressed in terms of the integration variable $u$ and the parameters $v, v^\prime, \theta$ through the linear transformation:
\begin{align}
q &= u\cos\theta + \frac{v+v'}{2}\sin\theta, \label{eq:q_transform} \\
p &= \frac{v+v'}{2}\cos\theta - u\sin\theta. \label{eq:p_transform}
\end{align}

Similarly, for the effect matrix \(E\), the matrix element is given by:
\begin{equation}
\langle v,\theta | E | v',\theta \rangle = \int_{-\infty}^{\infty} du \, W_E(q,p) \, e^{i u (v - v')},
\label{eq:matrix_element_E_integral}
\end{equation}
Substituting $W_E(q,p)$ from Eq.~(\ref{eq:Wigner_E}) and the same coordinate transformations, we obtain the corresponding matrix elements of \(E\).
In the framework of quantum trajectory theory and the past quantum state, the explicit form of the trace operation $\operatorname{Tr}(\hat{\Omega}_{m_2} \rho \hat{\Omega}_{m_2}^\dagger E)$ is obtained by expanding the operators in a complete basis of the atomic spin, for instance the rotated basis $|v,\theta\rangle $.  

Let $\hat{x}_A(\theta)$ denote the atomic spin projection operator along direction $\theta$, with eigenstates $|v, \theta\rangle$ satisfying the completeness relation: 
\begin{equation}
\int \mathrm{d}v \, |v,\theta\rangle\langle v,\theta| = \hat{I}_A .
\end{equation}
Expanding the trace in this basis gives  
\begin{equation}
\operatorname{Tr}
\left(
\hat{\Omega}_{m_2}\rho
\hat{\Omega}_{m_2}^{\dagger}E
\right)
=
\mathcal{N}
\iint \mathrm{d}v\,\mathrm{d}v_1\,
\psi(m_2-\widetilde{\kappa}_2v)
\psi^{*}(m_2-\widetilde{\kappa}_2v_1)
\langle v,\theta|\rho|v_1,\theta\rangle
\langle v_1,\theta|E|v,\theta\rangle .
\end{equation}
where $\mathcal{N}$ is the normalization coefficient.
\begin{figure*}[t]
	\centering
	\includegraphics[width=5in]{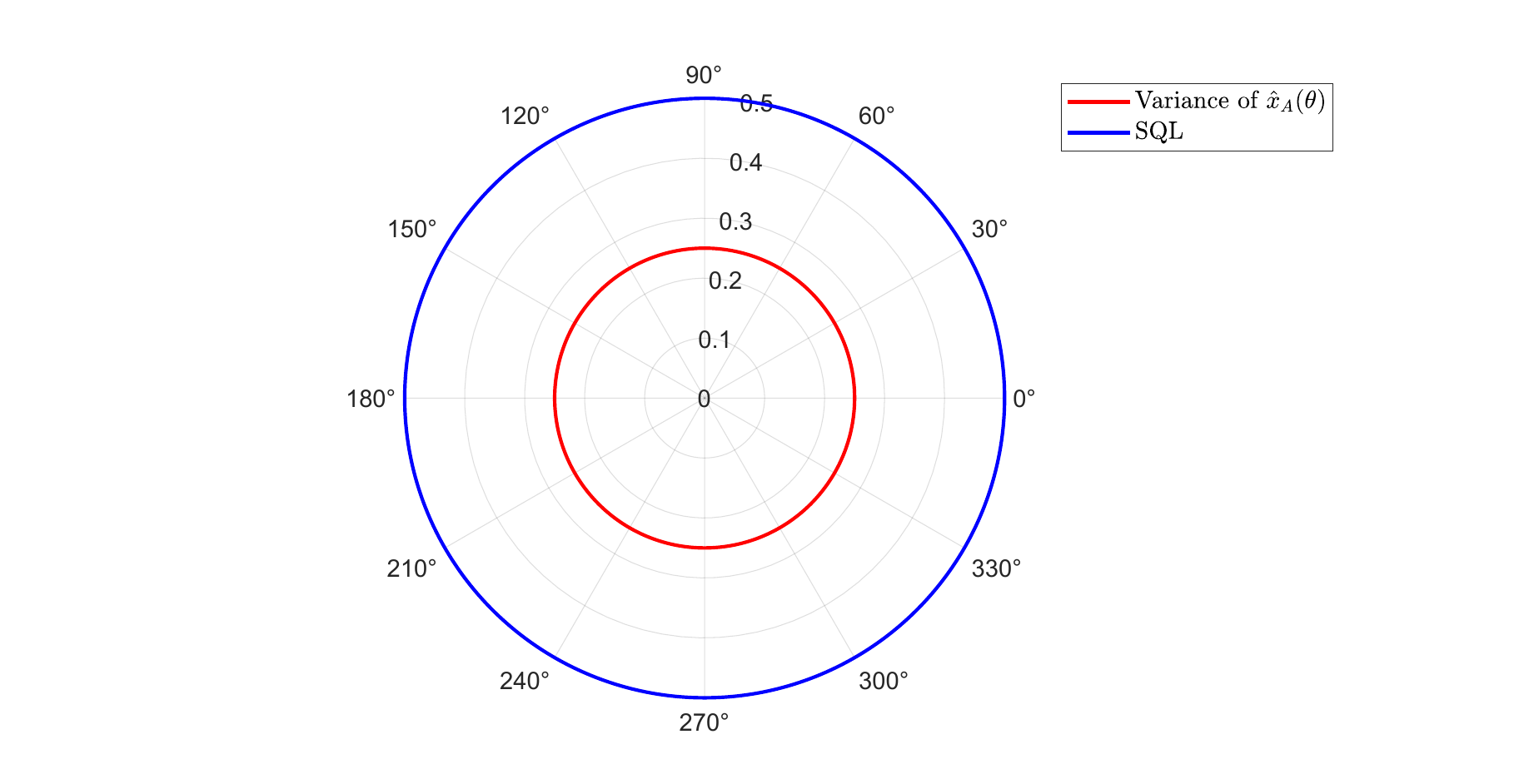}	
	\caption{{\bf Spin-noise variance below the SQL in all directions in theory.} 
		{\bf }{The blue line shows the variance of 0.5 at the standard quantum limit, while the red line shows the theoretically achievable variance of 0.25, reaching the 3 dB theoretical limit in all directions.} 
		}	
	\label{Fig:S1}			
\end{figure*}
The results of the calculated variance is 
\begin{equation}
\begin{aligned}
{\rm Var}_{m_2}(\sigma_1,\sigma_2,\sigma_3,\sigma_4,\theta,\widetilde{\kappa}_2)
= \frac{1}{
2 \Big[ \widetilde{\kappa}_2^2
(\sigma_1^2 + \sigma_3^2)\cos^2\theta
+ (\sigma_2^2 + \sigma_4^2)\big(\sin^2\theta\, \widetilde{\kappa}_2^2 + 8\sigma_1^2 + 8\sigma_3^2\big)
\Big]
}
\\
\times \Bigg[
2\cos^4\theta\, \widetilde{\kappa}_2^4 \sigma_1^2 \sigma_3^2
+ 2\widetilde{\kappa}_2^2 \Big(
\widetilde{\kappa}_2^2(\sigma_1^2 \sigma_4^2 + \sigma_2^2 \sigma_3^2)\sin^2\theta
+ (8\sigma_2^2\sigma_3^2 + 8\sigma_3^2\sigma_4^2 + \tfrac{1}{2})
\sigma_1^2 + \tfrac{\sigma_3^2}{2} \Big) \cos^2\theta
\\
+ 2\Big(
\sin^2\theta\, \widetilde{\kappa}_2^2 \sigma_2^2 \sigma_4^2
+ \tfrac{\sigma_2^2}{2} + \tfrac{\sigma_4^2}{2}
\Big)
\big(\sin^2\theta\, \widetilde{\kappa}_2^2 + 8\sigma_1^2 + 8\sigma_3^2\big)
\Bigg].
\end{aligned}
\end{equation}
In a fully symmetric case, $\sigma_1^2=\sigma_2^2=\sigma_3^2=\sigma_4^2= 0.5$, the result reduces to $\frac{1}{4}\widetilde{\kappa}_2^2  + \frac{1}{2}$. The variance of the verification outcome $m_2$ contains both photon shot noise and the atomic contribution. After subtracting the photon shot-noise term and normalizing by the measurement strength $\widetilde{\kappa}_2^2$, the inferred atomic variance reaches 0.25, which is 3 dB below the SQL of 0.5.


\section*{S4\quad Experimental imperfection and theoretical corrections}

In the ideal case of Sec. S2,  the forward state $\rho$ and the backward effect matrix $E$ are conditioned on a complete measurement record, such that all measurement back-action is consistently included in the quantum trajectory. In the experiment, however, two non-ideal effects must be taken into account. First, the second measurement stage (verification stage) itself has back-action noise due to the intertwining and alternating $x$ and $p$ measurements. Second, finite gap times are used to separate the first, second, and third measurement stages in order to remove the correlations caused by the LIA itself. During these gap intervals, the atoms are still affected by the probe back-action, but no measurement outcomes are recorded. Therefore, the corresponding evolution is only partially monitored, and the ideal quantum-trajectory description must be corrected.

To make the theory quantitatively comparable with the experiment, we incorporate these imperfections as corrections to both $\rho$ and $E$. Operationally, this amounts to discarding the stochastic measurement term during the gap intervals and in the verification stage, so that the corresponding evolution is described by an unconditional master equation. The corrected evolution in the unobserved intervals is written as
\begin{equation}\label{eq:s4_sme1}
    d \rho = (\mathcal{L}_{qba}+ \mathcal{L}_{damp})\rho \, d t,
    \qquad
    d E = (\mathcal{L}_{qba}^\dagger+ \mathcal{L}^\dagger_{damp})E \, d t,
\end{equation}
For Gaussian states, the effect of Eq.~\eqref{eq:s4_sme1} can be captured at the level of second moments. The forward variances obey
\begin{equation}\label{eq:s4_var}
\frac{d\sigma_x^2}{dt}= -2\Gamma_{\mathrm{tot}}\sigma_x^2 + 2\Gamma_{\mathrm{tot}}V_0 + \frac{\kappa_p^2}{2},
\qquad
\frac{d\sigma_p^2}{dt}= -2\Gamma_{\mathrm{tot}}\sigma_p^2 + 2\Gamma_{\mathrm{tot}}V_0 + \frac{\kappa_x^2}{2}.
\end{equation}
The first term in each equation describes the relaxation of the collective spin fluctuations at the total damping rate $\Gamma_{\mathrm{tot}}$, while the second term drives the system towards the unconditional steady-state variance $V_0$. The last term represents quantum back-action noise. Importantly, because no measurement record is used in these intervals, the nonlinear conditioning terms proportional to $\eta \kappa^2$ are absent.

Similarly, the backward-evolving effect matrix is broadened by the same unobserved back-action, and the corresponding retrodictive variances satisfy
\begin{equation}\label{eq:s4_ret}
\frac{d\gamma_x^2}{dt}= 2\Gamma_{\mathrm{tot}}\gamma_x^2 + 2\Gamma_{\mathrm{tot}}V_0 + \frac{\kappa_p^2}{2},
\qquad
\frac{d\gamma_p^2}{dt}= 2\Gamma_{\mathrm{tot}}\gamma_p^2 + 2\Gamma_{\mathrm{tot}}V_0 + \frac{\kappa_x^2}{2}.
\end{equation}
These equations have the same diffusion terms as the forward evolution, but with the sign of the drift reversed. Physically, Eq.~\eqref{eq:s4_ret} describes how the information carried by the later measurement record is degraded when propagated backward through time intervals in which back-action is present but no outcomes are registered. Considering the experimental imperfections, the ideal squeezing will degrade by about $0.27$ dB.

Finally, we note that the dominant decoherence mechanisms in the experiment are the atom-atom collisions, atom-wall collisions and residual magnetic field inhomogeneities, which give a damping rate of $\Gamma_c = 2\pi \times 60.6\:{\rm Hz}$. To maintain a steady state high spin polarization, a weak resonant pump beam continuously repopulates the atoms to the ground state of the spin oscillator \cite{duan2025concurrent}. In the dynamic equilibrium, the measured spin variance is $9\%$ higher than that of the calibrated ground state. Since the pumping process itself introduces additional decoherence via photon scattering, the total damping rate including pumping is $\Gamma_{\mathrm{tot}} = 2\pi \times331\:\mathrm{Hz}$.

\section*{S5\quad Effect of the probe duty cycle}

In this section, we show that the duty cycle of the probe does not affect the effective measurement strengths of the two rotating-frame quadratures,
provided that the average probe power is kept constant. In the laboratory frame, the spin quadrature coupled to the probe light is
\begin{equation}
\hat{p}_{\mathrm{A}}^{\prime}(t)
=
-\hat{x}_{\mathrm{A}}\sin(\Omega_L t)
+
\hat{p}_{\mathrm{A}}\cos(\Omega_L t).
\label{eq:duty_rotating_quadrature}
\end{equation}
For an infinitesimal measurement interval, the measurement strength is
proportional to $\kappa^2(t)$ whose temporal form is determined by the duty cycle. Projecting the measured quadrature onto the
orthogonal rotating-frame quadratures and integrating over the Larmor period $T_L=2\pi/\Omega_L$ gives the effective measurement strengths
\begin{equation}
\kappa_x^2
=
\frac{1}{T_L}
\int_0^{T_L}
\kappa^2(t)\sin^2(\Omega_L t)\,\mathrm{d}t,
\label{eq:duty_kappa_x}
\end{equation}
and
\begin{equation}
\kappa_p^2
=
\frac{1}{T_L}
\int_0^{T_L}
\kappa^2(t)\cos^2(\Omega_L t)\,\mathrm{d}t.
\label{eq:duty_kappa_p}
\end{equation}

For the stroboscopic modulation used in the experiment, the probe intensity
has fourfold temporal symmetry,
\begin{equation}
\kappa^2\left(t+\frac{T_L}{4}\right)
=
\kappa^2(t).
\label{eq:duty_fourfold_symmetry}
\end{equation}
Using
\begin{equation}
\sin^2\left(\Omega_L t+\frac{\pi}{2}\right)
=
\cos^2(\Omega_L t),
\end{equation}
Eqs.~\eqref{eq:duty_kappa_x} and \eqref{eq:duty_kappa_p} give
\begin{equation}
\kappa_x^2=\kappa_p^2.
\label{eq:duty_equal_strength}
\end{equation}
Furthermore, since
\begin{equation}
\sin^2(\Omega_L t)+\cos^2(\Omega_L t)=1,
\end{equation}
their sum is
\begin{equation}
\kappa_x^2+\kappa_p^2
=
\frac{1}{T_L}
\int_0^{T_L}
\kappa^2(t)\,\mathrm{d}t
\equiv
\overline{\kappa^2},
\label{eq:duty_total_strength}
\end{equation}
where $\overline{\kappa^2}$ is the measurement strength averaged over one
Larmor period. Therefore,
\begin{equation}
\kappa_x^2
=
\kappa_p^2
=
\frac{\overline{\kappa^2}}{2}.
\label{eq:duty_final_strength}
\end{equation}

When the duty cycle is increased, each probe pulse covers a wider range of phases in a Larmor period, so the instantaneous measured quadrature is no longer exactly $\hat{x}_{\mathrm{A}}$ or $\hat{p}_{\mathrm{A}}$. However, phase-sensitive demodulation separates the measurement record into the two orthogonal rotating-frame quadratures. Owing to the fourfold symmetry of the pulse sequence, the integrated measurement strengths of $\hat{x}_{\mathrm{A}}$ and $\hat{p}_{\mathrm{A}}$ remain equal. If the average probe power is kept constant, both $\kappa_x^2$ and $\kappa_p^2$ are therefore independent of the duty cycle, explaining the experimentally observed duty-cycle-insensitive squeezing.

\section*{S6\quad Calibration of the standard quantum limit}

\begin{figure*}[t]
	\centering
	\includegraphics[width=7in]{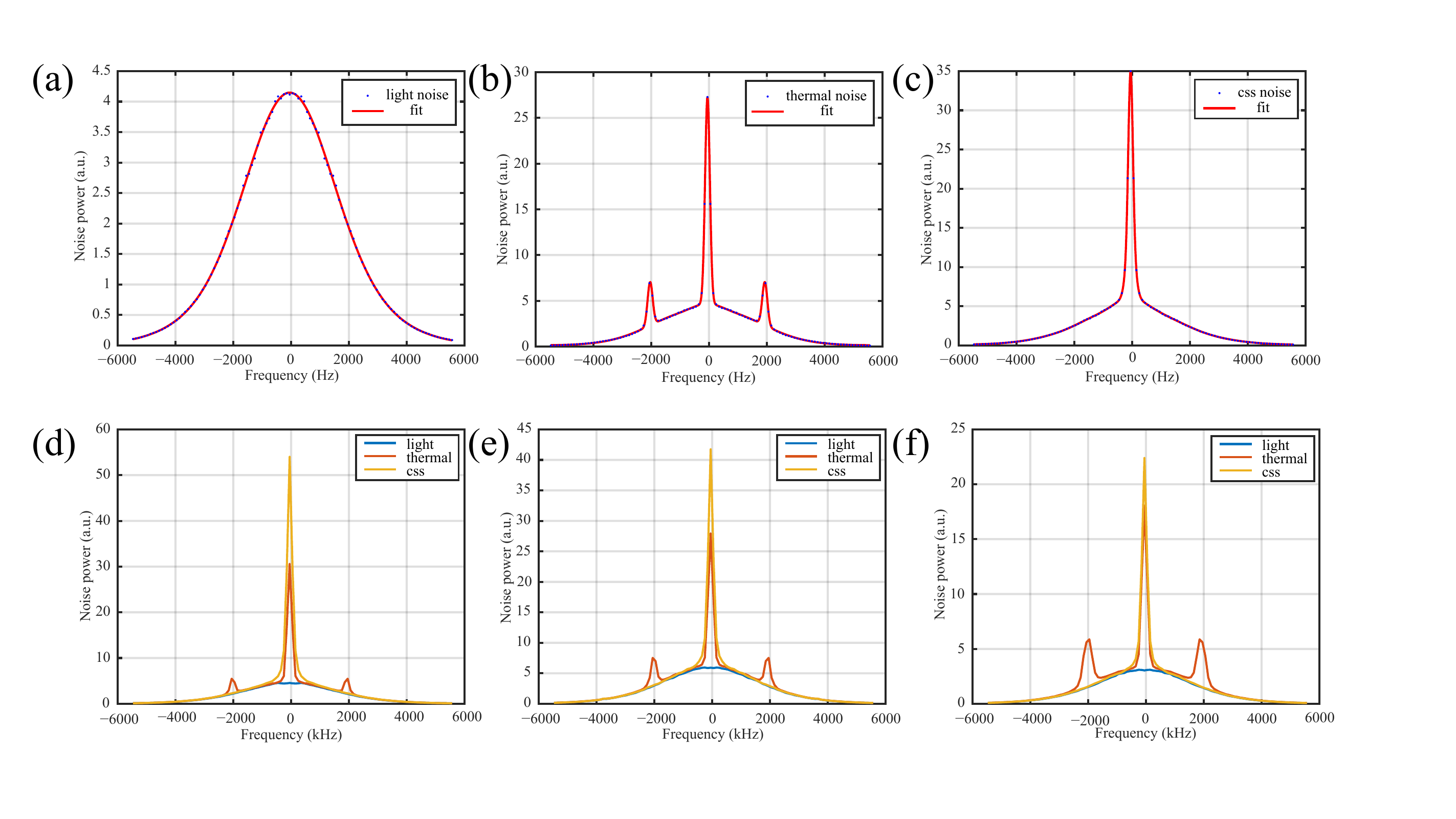}	
	\caption{{\bf Noise frequency spectrum } 
		{\bf (a)}{ Photon shot noise spectrum of light. The spectrum exhibits the typical characteristics of shot noise, where the spectral lines correspond to the response curve of the LIA, measured with the LIA low-pass filter bandwidth configured to 2 kHz.} 
		{\bf (b)}{ Thermal state noise spectrum. This spectrum shows two small peaks at 2 kHz, which are attributed to atomic populations in the F=1 state.}
		{\bf (c)}{ Coherent spin state spectrum. The narrow feature at the center of the spectrum corresponds to the coherence time of the atomic spins. The broader structure is a result of the  LIA's response time.}
		{\bf (d)-(f)}{ Comparison of frequency spectra for probe light at blue detuning of 4 GHz, 4.5 GHz, and 5 GHz. As the detuning of the probe light increases, the thermal state noise spectrum shows increased contributions from atoms in the F=1 state. This is manifested as the progressively larger peaks around ±2 kHz.}}
	\label{Fig:2}			
\end{figure*}

In spin-squeezing experiments, calibration of the standard quantum limit (SQL) is of particular importance. In practical measurements, the detected signal contains contributions from the detector’s background electronic noise, photon shot noise, and atomic noise arising from atom-light interactions. The electronic noise is negligible in the experiment with the lock-in amplifier (LIA) \cite{duan2025concurrent}. Keeping the demodulation frequency of the LIA fixed, we measure the photon shot noise by varying the bias magnetic field to shift the Larmor frequency out of the  frequency band of the LIA demodulation and of the atomic noise. Unpolarized atoms in a thermal state are insensitive to external magnetic field fluctuations and other environmental perturbations, providing excellent robustness, and can therefore be used to calibrate the SQL. In the thermal state, the atoms are equally distributed among the eight Zeeman sublevels of F=1 and F=2, leading to an SQL that corresponds to 0.8 times the thermal atomic noise level \cite{bao2020spin}:

\begin{equation}\label{eq:5}
\begin{aligned}
\mathrm{SQL} = 0.8\epsilon_1  \left( \mathrm{Var}(\hat{S}_{y}^{thermal}) - \mathrm{Var}(\hat{S}_y^{light}) \right).
\end{aligned}
\end{equation}
The correction factor $\epsilon_1$ accounts for the correction of atoms on ground states F=1. Because of the difference in the Landé g-factors of the ground states F=1 and F=2, the Larmor frequency of atoms in F=1 differs from that of atoms in F=2 by approximately 2 kHz. This discrepancy can be corrected using the spectral response, as illustrated in Fig.\ref{Fig:2}.

\section*{S7\quad Data analysis}

The detected signal is demodulated at the Larmor frequency using LIA to obtain the output result:$\bold{R} = \left\{\bold{R_x},\bold{R_y}\right\}$. In the experiment, the probe light is stroboscopically modulated at four times the Larmor frequency. By adjusting the demodulation phase, the information of $\hat{x}_A$ and $\hat{p}_A$ can be separated, corresponding to $\bold{R_x}$ and $\bold{R_y}$, respectively. Therefore, we can simultaneously measure the quadratures $\hat{x}_A$ and $\hat{p}_A$, and evaluate their squeezing degrees separately.

To evaluate the degree of squeezing, we make use of the correlations between different segments of the measurement record. The data are divided into three time intervals: the heralding, verification, and postselection intervals, with their corresponding effective measurement outcomes denoted by $m_1$, $m_2$, and $m_3$, respectively. The heralding and postselection records provide the earlier and later information used to condition the verification outcome $m_2$. Based on the temporal correlations between $m_1$, $m_3$ and the verification segment $m_2$, a time-mode processing method is employed to fully extract information from the data \cite{krauter2013deterministic}. When only the forward conditioning is applied, the minimum variance of $m_2$ conditioned on $m_1$ is
\begin{equation}\label{eq:6_1}
\begin{aligned}
\mathrm{Var}\left(m_2\mid m_1\right)&=\min_{\alpha} \left[\mathrm{Var}\left(m_2-\alpha m_1\right)\right]\\
&=\mathrm{Var}\left(m_2\right)-\frac{\mathrm{Cov}^2\left(m_2,m_1\right)}{\mathrm{Var}\left(m_1\right)}.
\end{aligned}
\end{equation}
Employing the past quantum state theory, which simultaneously accounts for both prediction and retrodiction, the minimum variance of $m_2$ conditioned on $m_1$ and $m_3$ is given by:

\begin{equation}\label{eq:7}
\begin{aligned}
\mathrm{Var}(m_2 \mid m_1, m_3) &= \min_{\alpha, \beta} \left[\mathrm{Var}(m_2 - \alpha m_1 - \beta m_3)\right] \\
&= \min_{\alpha, \beta} \bigl[\mathrm{Var}(m_2) + \alpha^2 \mathrm{Var}(m_1) + \beta^2 \mathrm{Var}(m_3) \\
&\quad - 2\alpha \mathrm{Cov}(m_2, m_1) - 2\beta \mathrm{Cov}(m_2, m_3) + 2\alpha \beta \mathrm{Cov}(m_1, m_3)\bigr]\\
&= \mathrm{Cov}_{22} + \alpha^2 \mathrm{Cov}_{11} + \beta^2 \mathrm{Cov}_{33} - 2\alpha \mathrm{Cov}_{21} - 2\beta \mathrm{Cov}_{23} + 2\alpha\beta  \mathrm{Cov}_{13},
\end{aligned}
\end{equation}
where $\mathrm{Cov}_{uv}$ is the covariance between $m_u$ and $m_v$, and when $u=v$, $\mathrm{Cov}_{uu}=\mathrm{Var}(m_u)$, and 
\begin{equation}\label{eq:8}
\begin{aligned}
\alpha &= \frac{\mathrm{Cov}_{21}\mathrm{Cov}_{33} - \mathrm{Cov}_{23}\mathrm{Cov}_{13}}{\mathrm{Cov}_{11}\mathrm{Cov}_{33} - \mathrm{Cov}_{13}^{2}}, \\
\beta &= \frac{\mathrm{Cov}_{23}\mathrm{Cov}_{11} - \mathrm{Cov}_{21}\mathrm{Cov}_{13}}{\mathrm{Cov}_{11}\mathrm{Cov}_{33} - \mathrm{Cov}_{13}^{2}}.
\end{aligned}
\end{equation}

\section*{S8\quad Wineland criterion}

While there are various definitions of the squeezing parameter, we adopt the one proposed by Wineland, which is defined as the ratio between the phase resolution of the squeezed state and that of the spin coherent state.
\begin{equation}\label{eq:8_1}
\begin{aligned}
\xi_{R}^{2}=\frac{(\Delta\phi)^{2}}{(\Delta\phi)_{\text{css}}^{2}}=\frac{\epsilon_2 \mathrm{Var}_{min}(m_2) }{0.8\epsilon_1  \left( \mathrm{Var}(\hat{S}_{y}^{thermal}) - \mathrm{Var}((\hat{S}_y^{light})) \right)}.
\end{aligned}
\end{equation}
Here, $\Delta\phi$ and $\Delta\phi_{\text{css}}$ are the angular resolutions of the spin-squeezed state and the coherent spin state (CSS), respectively, with the latter at a polarization of unity corresponding to the projection noise limit (PNL).

In the steady-state experiment, the collective spin component remains unchanged under continuous optical pumping, and the degree of polarization is maintained at 96.91\%. The corresponding reduction of the spin length is taken into account by introducing a correction factor $\epsilon_2$. In this work, we determine $\epsilon_2=1.06$, which is applied to calibrate the effective spin size in subsequent analyses.
\bibliography{scisibib}